\documentclass[runningheads,a4paper]{llncs}

\usepackage{caption}
\usepackage{subcaption}
\usepackage{amsmath}
\usepackage{amssymb}
\usepackage{listings}
\usepackage{multirow}
\usepackage{multicol}
\usepackage{fancyvrb}
\usepackage{xcolor}

\usepackage{xspace}
\usepackage{tikz}
\usepackage{cite}
\usepackage{pgfplots}
\usepackage{hyperref}

\newcommand{\figref}[1]{Fig.~\ref{Fi:#1}}
\newcommand{\figrefs}[2]{Figs.~\ref{Fi:#1} and~\ref{Fi:#2}}

\renewcommand{\eqref}[1]{Eqn.~(\ref{Eq:#1})}

\newcommand{\tableref}[1]{Tab.~\ref{Ta:#1}}

\newcommand{\sectref}[1]{Section~\ref{Se:#1}}

\newcommand{\inroundb}[1]{\ensuremath{ \left( #1 \right)}}
\newcommand{\inparan}[1] {\ensuremath{ \{ #1 \}} }
\newcommand{\vbars}[1]{\ensuremath{\left\vert #1 \right\vert}}

\newcommand{\SSZ}[1]{\scriptsize{#1}}

\newcommand{\TT}[1]{\texttt{#1}}

\newcommand{\MC}[1]{\ensuremath{\mathcal{#1}}}
\newcommand{\BF}[1]{\textbf{#1}}
\newcommand{\EM}[1]{\emph{#1}}
\newcommand{\Omit}[1]{}

\newcommand{\Nat}{\ensuremath{\mathbb{N}}}

\newcommand{\pblib}{\textsc{PBLib}\xspace}
\newcommand{\pedigree}{\textsf{pedigree}\xspace}
\newcommand{\sorter}{\textsf{sorter}\xspace}
\newcommand{\swc}{\textsf{swc}\xspace}
\newcommand{\adder}{\textsf{adder}\xspace}
\newcommand{\bdd}{\textsf{bdd}\xspace}
\newcommand{\binmerge}{\textsf{bin-merger}\xspace}
\newcommand{\gte}{\textsf{gte}\xspace}
\newcommand{\watchdog}{\textsf{watchdog}\xspace}

\newcommand{\minisat}{\textsc{Minisat 2.2.0}\xspace}

\newcommand{\true}{$true$\xspace}
\newcommand{\false} {$false$\xspace}

\definecolor{highlighter}{rgb}{0.85,0.85,0.85}

\makeatletter
\newbox\sf@box
{\def\sf@one{#1}%
\def\sf@two{#2}%
\setbox\sf@box\hbox
\bgroup}%
{ \egroup
\ifx\@empty\sf@two\@empty\relax
\def\sf@two{\@empty}
\fi
\ifx\@empty\sf@one\@empty\relax
\subfloat[\sf@two]{\box\sf@box}%
\else
\subfloat[\sf@one][\sf@two]{\box\sf@box}%
\fi}
\makeatother

\usepackage{algorithm}
\usepackage{algorithmic}

\newif\ifhenabled

\urldef{\mailsa}\path|{saurabh.joshi,ruben.martins}@cs.ox.ac.uk|
\urldef{\mailsb}\path|vasco.manquinho@inesc-id.pt|

\begin{document}

\mainmatter

\title{Generalized Totalizer Encoding for Pseudo-Boolean Constraints}


\author{Saurabh Joshi\inst{1} \and Ruben Martins\inst{1} \and Vasco Manquinho\inst{2}}


\institute{Department of Computer Science, University of Oxford, UK \\
	\mailsa
	\and
	INESC-ID / Instituto Superior T\'ecnico, Universidade de Lisboa, Portugal \\
	\mailsb
 }

\toctitle{}
\tocauthor{}
\maketitle
\begin{abstract}
Pseudo-Boolean constraints, also known as 0-1 Integer Linear Constraints, are used to model many real-world problems.
%
A common approach to solve these constraints is to 
encode them into a SAT formula. 
The runtime of the SAT solver on such formula is sensitive to the manner in which the 
given pseudo-Boolean constraints are encoded.
In this paper, we propose generalized Totalizer encoding (GTE), which is an arc-consistency preserving extension of the Totalizer encoding to pseudo-Boolean constraints.
Unlike some other encodings, the number of auxiliary variables required for GTE does not depend on the magnitudes of the coefficients. Instead, it depends on the number of distinct combinations of these coefficients.
We show the superiority of GTE with respect to other encodings when large pseudo-Boolean constraints have low number of distinct coefficients.
%
Our experimental results also show that 
GTE remains competitive even when the pseudo-Boolean constraints do not have this characteristic.

\end{abstract}

\section{Introduction \label{Se:introduction}}
Pseudo-Boolean constraints (PBCs) or 0-1 Integer Linear constraints have been used to
model a
plethora of real world problems such as computational biology~\cite{pedigrees,pbcbio}, 
upgradeability problems~\cite{argelich10,packup,ignatiev-icse14}, resource allocation~\cite{pbcvm}, scheduling~\cite{pbcsched} and automated test pattern generation~\cite{marques-silva-asp-dac98}.
Due to its importance and a plethora of applications, a lot of research has been done to efficiently solve PBCs.
One of the popular approaches is to convert PBCs into a SAT formula~\cite{een-jsat06,npsolver,watchdog} 
thus making them amenable to off-the-shelf SAT solvers.
We start by formally introducing PBC, followed by a discussion on how to convert a PBC into a SAT formula.


A PBC is defined over a finite set of Boolean variables $x_1,\dots,x_n$ which
can be assigned a value 0 (\false) or 1 (\true). A literal $l_i$ is either a Boolean
variable $x_i$ (positive literal) or its negation $\neg x_i$ (negative literal).
A positive (resp. negative) literal $l_i$ is said to be assigned 1 if and only if
the corresponding variable $x_i$ is assigned 1 (resp. 0). Without a loss of generality,
PBC can be defined as a linear inequality of the following normal form:

\begin{align}
\sum w_il_i \leq k
\end{align}

Here, $w_i \in \Nat^+$ are called coefficients or weights, $l_i$ are input literals and
$k \in \Nat^+$ is called the bound. Linear inequalities in other forms (e.g. other
inequality, equalities or negative coefficients) can be converted into
this normal form in linear time~\cite{barth96}. \EM{Cardinality constraint} is a special
case of PBC when all the weights have the value $1$. Many different encodings have been proposed
to encode cardinality constraints\cite{bailleux-cp03,seq,asin-constraints11,totalizer-ictai13}. 
%
Linear pseudo-Boolean solving (PBS) is a generalization of the SAT formulation 
where constraints are not restricted to clauses and can be PBCs. 
A related problem to PBS is the linear pseudo-Boolean 
optimization (PBO) problem, where all the constraints must be satisfied and the 
value of a linear cost function is optimized. PBO usually requires an iterative 
algorithm which solves a PBS in every iteration~\cite{een-jsat06,npsolver,sat4j,wbo}. 
Considering that the focus of the paper is on encodings rather than algorithms, 
we restrict ourselves to the decision problem (PBS).

This paper makes the following contributions.
\begin{itemize}
\item We propose an arc-consistency~\cite{gac} preserving extension of Totalizer
encoding~\cite{bailleux-cp03} called Generalized Totalizer encoding (GTE) in
\sectref{gentotal}.
\item We compare various PBC encoding schemes that were implemented in a common framework,
thus providing a fair comparison. After discussing related work in \sectref{related}, we show GTE as a promising encoding through its
competitive performance in \sectref{results}.
\end{itemize}
\section{Generalized Totalizer Encoding \label{Se:gentotal}}
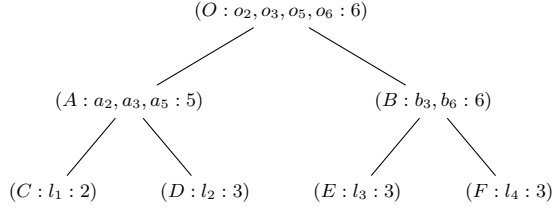
\begin{figure}[!t]
\centering
\begin{tikzpicture}[level/.style={sibling distance=50mm/#1},scale=0.8, every node/.style={scale=0.8}]
\Omit{\node (a) {$(A:o_1,o_2,o_3,o_4,o_5:5)$}
	child { node (b) {$(B:s^1 _1,s^1 _2:2)$ } 
		child { node (d) {$(D:l_1:1)$} } 
		child { node (e) {$(E:l_2:1)$} } }
	child { node (c) {$(C:s^2 _1, s^2 _2, s^2 _3:3)$} 
                child { node (g) { $(G: l_3:1)$}}
		child { node (f) {$(F:s^3 _1, s^3 _2:2)$} 
			child { node (h) {$(H:l_4:1)$}}
			child { node (i) {$(I:l_5:1)$}}} };}
\node (a) {$(O:o_2,o_3,o_5,o_6:6)$}
	child { node (b) {$(A:a_2,a_3,a_5:5)$ } 
		child { node (d) {$(C:l_1:2)$} } 
		child { node (e) {$(D:l_2:3)$} } }
	child { node (c) {$(B:b_3, b_6:6)$} 
                child { node (g) { $(E: l_3:3)$}}
		child { node (f) {$(F:l_4:3)$} }
 };
\end{tikzpicture}

\caption{Generalized Totalizer Encoding for $2l_1 + 3l_2 + 3l_3 + 3l_4 \leq 5$}
\label{Fi:gte}
\end{figure}

The Totalizer encoding~\cite{bailleux-cp03} is an encoding to convert cardinality constraints into a SAT formula.
In this section, the generalized Totalizer encoding (GTE) to encode PBC
into SAT is presented.
GTE can be better visualized as a binary tree, as shown in \figref{gte}. With the exception of the leaves, every node
is represented as ($node\_name:node\_vars:node\_sum$). The $node\_sum$ for every node represents the maximum possible weighted sum of the subtree rooted at that node. For any node $A$,
a node variable $a_w$ represents a weighted sum $w$ of the underlying subtree. In other words, whenever the weighted sum of some of the input literals in the subtree becomes $w$, $a_w$ must be
set to $1$.
Note that for any node $A$, we would need one variable corresponding to
every distinct weighted sum that the input literals under $A$ can produce. Input literals are at the leaves,
represented as ($node\_name:literal\_name:literal\_weight$) with each of the terms being self
explanatory.

For any node $P$ with children $Q$ and $R$, to ensure that weighted sum is propagated from
$Q$ and $R$ to $P$, the following formula is built for $P$:
\begin{align}
\resizebox{.93\hsize}{!}{$
	\inroundb{
\bigwedge _{ 
\begin{scriptsize}
\begin{array}{c}
q_{w_1} \in Q.node\_vars \\
r_{w_2} \in R.node\_vars \\
w_3 = w_1 + w_2\\
p_{w_3} \in P.node\_vars \\
\end{array}
\end{scriptsize} }
\inroundb{
\neg q_{w_1} \vee \neg r_{w_2} \vee p_{w_3}}
}
\wedge
\inroundb{
\bigwedge _{
\begin{scriptsize}
\begin{array}{c}
s_w \in \inroundb{Q.node\_vars \cup R.node\_vars} \\
w=w'\\
p_{w'} \in P.node\_vars
\end{array}
\end{scriptsize}
} \inroundb{\neg s_w \vee p_{w'}}
}
\label{Eq:gte}
$}
\end{align}

The left part of \eqref{gte} ensures that, if node $Q$ has witnessed a weighted sum of
$w_1$ and $R$ has witnessed a weighted sum of $w_2$, then $P$ must be considered to have
witnessed the weighted sum of $w_3=w_1+w_2$. The right part of \eqref{gte} just takes
care of the boundary condition where weighted sums from $Q$ and $R$  are propagated to $P$ without
combining it with their siblings. This represents that $Q$ (resp. $R$) has witnessed a weighted
sum of $w$ but $R$ (resp. $Q$) may not have witnessed any positive weighted sum.

Note that node $O$ in \figref{gte} does not have variables for
the weighted sums larger than $6$. Once the weighted sum goes above the threshold of $k$,
we represent it with $k+1$. Since all the weighted sums above $k$ would result in the
constraint being not satisfied, it is sound to represent all such sums as $k+1$.
This is in some sense a generalization of $k$-simplification described
in \cite{buttner-icaps05,qmaxsat-jsat12}. For $k$-simplification, $w_3$ in \eqref{gte} would change to $w_3 =
min(w_1+w_2,k+1)$.

Finally, to enforce that the weighted sum does not exceed the given threshold $k$, we add
the following constraint at the root node $O$ : 
\begin{align}
\neg o_{k+1}
\end{align}

\noindent \BF{Encoding properties: }
Let $A_{I_w}$ represent the multiset
of weights of all the input literals in the subtree rooted at node $A$. For any given multiset $S$
of weights, let $Weight(S) = \sum _{e \in S} e$. For a given multiset $S$, let $unique(S)$
denote the set with all the multiplicity removed from $S$.  
Let $\vbars{S}$ denote the cardinality of the set $S$. Hence,
the total number of node variables required at node $A$ is:

\begin{align}
\vbars{
unique\inroundb{\inparan{Weight(S)| S \subseteq A_{I_w} \wedge S \neq \emptyset}} }
\label{Eq:gtesize}
\end{align}

Note that unlike some other encodings~\cite{swc,watchdog} the number of auxiliary variables
required for GTE does not depend on the magnitudes of the weights. Instead, it depends on
how many unique weighted sums can be generated. Thus, we claim that for pseudo-Boolean
constraints where the distinct weighted sum combinations are low, GTE should perform
better. We corroborate our claim in \sectref{results} through experiments.

Nevertheless, in the worst case, GTE can generate exponentially many auxiliary variables and
clauses. For example, if the weights of input literals $l_1,\dots,l_n$ are respectively
$2^0,\dots,2^{n-1}$, then every possible weighted sum combination would be unique.
In this case, GTE would generate exponentially many auxiliary variables. Since every
variable is used in at least one clause, it will also generate exponentially many clauses.

Though GTE does not depend on the magnitudes of the weights, one can use the magnitude of the 
largest weight to categorize a class of PBCs for which GTE is guaranteed to be of polynomial size.
If there are $n$ input literals and the largest weight is a polynomial $P(n)$, then GTE is guaranteed 
to produce a polynomial size formula. If the largest weight is $P(n)$, then the total number of distinct
weight combinations (\eqref{gtesize}) is bounded by $nP(n)$, resulting in a polynomial size formula.

The best case for GTE occurs when all of the weights are equal, in which case the number of
auxiliary variables and clauses is, respectively, $\MC{O}(n\,log_2n)$ and $\MC{O}(n^2)$. 
Notice that for this best case with $k$-simplification, we have $\MC{O}(nk)$ variables and clauses,
since it will behave exactly as the Totalizer encoding~\cite{bailleux-cp03}.

Note also that the generalized arc consistency (GAC)~\cite{gac} property of Totalizer encoding 
holds for GTE as well.
GAC is a property of an encoding which allows the solver to
         infer maximal possible information through propagation, thus
	          helping the solver to prune the search space earlier. 
The original proof~\cite{bailleux-cp03} makes an inductive argument using the left subtree 
and the right subtree of a node. It makes use of the fact that, if there are $q$ input variables
set to $1$ in the left child $Q$ and $r$ input variables are set to $1$ in the right
child $R$, then the encoding ensures that in the parent node $P$, the variable $p_{q+r}$ is
set to $1$. 
Similarly, GTE ensures that if the left child $Q$ contributes $w_1$ to the weighted sum
($q_{w_1}$ is set to $1$) and the right child $R$ contributes $w_2$ to the weighted sum
($r_{w_2}$ is set to $1$), then the parent node $P$ registers the weighted sum to be
at least $w_3=w_2+w_1$ ($p_{w_3}$ is set to $1$). Hence, the GAC proof still holds for
GTE. 
\Omit{
An alternative proof sketch for GAC is as follows. Let set $T$ be the set of input literals set to $1$
and the weighted sum over $T$ is exactly $k$. Any other literal not belonging to
$T$ can not be set to $1$ as otherwise, the total weight would exceed $k$, thus conflicting
with $\neg o_{k+1}$. Starting with the clause $\neg o_{k+1}$, information is propagated 
downwards through $\inroundb{\neg q_{w_1} \vee \neg r_{w_2} \vee p_w}$, 
where at the top level $p_w = o_{k+1}$. 
Without the loss of generality, let $q_{w_1}$ represent a combination purely coming from $T$, therefore set to $1$. 
As a result, $r_{w_2}$ will be forced to be $0$. Finally, this propagation will reach the leaves forcing the
input literals not in $T$ to be $0$.
}

\section{Related Work\label{Se:related}}
The idea of encoding a PBC into a SAT formula is not new. One of the first such encoding is described in~\cite{adder,een-jsat06}
which uses binary adder circuit like formulation to compute the weighted sum and then 
compare it against the threshold $k$. This encoding creates $\mathcal O (n\, log_2 k)$
 auxiliary clauses, but it is not arc-consistent. 
Another approach to encode PBCs into SAT is to use sorting networks~\cite{een-jsat06}. 
This encoding produces $\mathcal O (N\, log_2^2 N$) auxiliary clauses, where N is bounded
by $\lceil log_2 w_1 \rceil + \ldots + \lceil log_2 w_n \rceil$. This encoding is also not arc-consistent for PBCs, 
but it preserves more implications than the adder encoding, and it maintains GAC for cardinality constraints.
%

The Watchdog encoding~\cite{watchdog} scheme uses the
Totalizer encoding, but in a completely different manner than GTE. It uses multiple
Totalizers, 
one for each bit of the binary representation of the weights. The Watchdog encoding 
was the first polynomial sized encoding that maintains GAC for PBCs and it only 
generates $\mathcal O(n^3 log_2n\,log_2w_{max})$ auxiliary clauses. 
Recently, the Watchdog encoding has been generalized to a more abstract framework 
with the Binary Merger encoding~\cite{binmerge}. Using a different translation of 
the components of the Watchdog encoding allows the Binary Merger encoding to 
further reduce the number of auxiliary clauses to $\mathcal O(n^2 log_2^2n\,log_2w_{max})$. 
The Binary Merger is also polynomial and maintains GAC.

Other encodings that maintain GAC can be exponential in the worst case scenario, such as
BDD based encodings~\cite{bailleux-jsat06,een-jsat06,abio-jair12}. These encodings share 
quite a lot of similarity to GTE, such as GAC and independence from 
the magnitude of the weight. One of the differences is that GTE 
always has a tree like structure amongst auxiliary variables and input literals. 
However, the crucial difference lies in the manner in which auxiliary variables are generated, and what they represent. 
In BDD based approaches, an auxiliary variable $D_i$ attempts to reason about the weighted sum of the input 
literals either $l_i,\dots, l_n$ or $l_1,\dots,l_i$. On the other hand, an auxiliary variable $a_w$ at a node $A$ in GTE 
attempts to only reason about the weighted sum of the input literals that are descendants of $A$. 
Therefore, two auxiliary variables in two disjoint subtrees in GTE are guaranteed to reason
about disjoint sets of input literals. We believe that such a localized reasoning could be
a cause of relatively better performance of GTE as reported in \sectref{results}.
It is worth noting that the worst case scenario for GTE, when weights are of the form $a^i$, where $a\geq 2$, would generate a polynomial size formula 
for BDD based approaches~\cite{bailleux-jsat06,een-jsat06,abio-jair12}.

As GTE generalizes the Totalizer encoding, the Sequential Weighted Counter (SWC)
encoding~\cite{swc} generalizes sequential encoding~\cite{seq} for PBCs. Like BDD based approaches and GTE,
SWC can be exponential in the worst case. 

\section{Implementation and Evaluation \label{Se:results}}

All experiments were performed on two AMD 6276 processors (2.3 GHz) 
running Fedora 18 with a timeout of 1,800 seconds and a memory limit of 16 GB. 
Similar resource limitations were used during the last pseudo-Boolean (PB) evaluation
of 2012\footnote{\label{Footnote:pb12}\url{http://www.cril.univ-artois.fr/PB12/}}.
%
For a fair comparison, we implemented GTE (\gte) in the \pblib~\cite{pblib} (version 1.2)
open source library which contains a plethora of encodings, namely,
Adder Networks (\adder)~\cite{adder,een-jsat06}, Sorting Networks 
(\sorter)~\cite{een-jsat06}, watchdog (\watchdog)~\cite{watchdog}, 
Binary Merger (\binmerge)~\cite{binmerge}, Sequential Weighted Counter 
(\swc)~\cite{swc}, and 
BDDs (\bdd)~\cite{abio-jair12}. A new encoding in \pblib
can be added by implementing \TT{encode} method of the base class \TT{Encoder}.
Thus, all the encodings mentioned above, including GTE, only differ in how \TT{encode}
is implemented while they share the rest of the whole environment.
\pblib provides parsing and normalization~\cite{een-jsat06} routines for PBC
and uses
\minisat~\cite{minisat} as a back-end SAT solver.
When the constraint to be encoded into CNF is a cardinality constraint, we use 
the default setting of \pblib that dynamically selects a cardinality encoding 
based on the number of auxiliary clauses. When the constraint to be encoded 
into CNF is a PBC, we specify one of the above encodings.
\Omit{
With this setup we can fairly evaluate the impact of different pseudo-Boolean 
encodings when solving pseudo-Boolean problems, since the pseudo-Boolean 
encoding is the only difference between the different runs.}
%
%

\vspace{2mm}
\noindent \BF{Benchmarks: }
Out of 
all 355 instances from the \textsf{DEC-SMALLINT-LIN} category in the 
last PB evaluation of 
2012 (PB'12),
we only considered those 214 instances\footnote{Available at \url{http://sat.inesc-id.pt/~ruben/benchmarks/pb12-subset.zip}} that
contain at least $1$ PBC.
%
We also consider an additional set of \pedigree benchmarks from computational 
biology~\cite{pedigrees}. These benchmarks were originally encoded in 
Maximum Satisfiability (MaxSAT) and were used in the last MaxSAT Evaluation of 
2014\footnote{\url{http://www.maxsat.udl.cat/14/}}. Any MaxSAT problem can be 
converted to a corresponding equivalent pseudo-Boolean problem~\cite{maxsat2pbo}.
We generate two pseudo-Boolean decision problems (one satisfiable, 
another unsatisfiable) from the optimization version of each of these benchmarks. 
The optimization function is transformed into a PBC with the value of the bound $k$ set
to a specific value. Let the optimum value for the optimization function be $k_{opt}$.
The satisfiable decision problem uses $k_{opt}$ as the value for the bound $k$, whereas the 
unsatisfiable decision problem uses $k_{opt}-1$ as the value for the bound $k$. 
Out of 200 generated 
instances\footnote{Available at \url{http://sat.inesc-id.pt/~ruben/benchmarks/pedigrees.zip}}, 
172 had at least $1$ PBC and were selected for further 
evaluation. 
\begin{table}[!t]
\caption{Characteristics of pseudo-Boolean benchmarks
}\label{Ta:properties}
\centering
\begin{tabular}{l|r|r|r|r|r|r}
\hline
Benchmark & \#PB & \#lits & $k$ & max $w_i$ & $\sum w_i$ & \#diff $w_i$\\
\hline
PB'12 & 164.31 & 32.25  & 27.94  & 12.55  & 167.14 & 6.72\\
\pedigree & 1.00 & 10,794.13 & 11,106.69 & 456.28  & 4,665,237.38 & 2.00\\
\hline
\end{tabular}
\end{table}
\begin{table}[!t]
\caption{Number of solved instances}\label{Ta:solved}
\centering
\begin{tabular}{ll|r|r|r|r|r|r|r}
\hline
Benchmark & Result & \sorter & \swc & \adder & \watchdog & \binmerge & \bdd & \gte\\
\hline
PB'12 & SAT & 72 & 74 & 73 & 79 & 79 & \textbf{81} & \textbf{81}\\
(214) & UNSAT & 74 & 77 & 83 & \textbf{85} & \textbf{85} & 84 & 84\\
\hline
\pedigree & SAT & 2 & 7 & 6 & 25 & 43 & 82 & \textbf{83}\\
(172) & UNSAT & 0 & 7 & 6 & 23 & 35 & 72 & \textbf{75}\\
\hline
Total & SAT/UNSAT & 146 & 165 & 172 & 212 & 242 & 319 & \textbf{323}\\
\hline
\end{tabular}
\end{table}
\pgfplotscreateplotcyclelist{myplotstylelist}
{
blue,mark=*\\
green,mark=o\\
dashed,brown, mark=star\\
densely dashdotted,cyan,mark=square\\
lime,mark=pentagon \\
dotted,orange,mark=diamond\\
black,mark=triangle\\
}

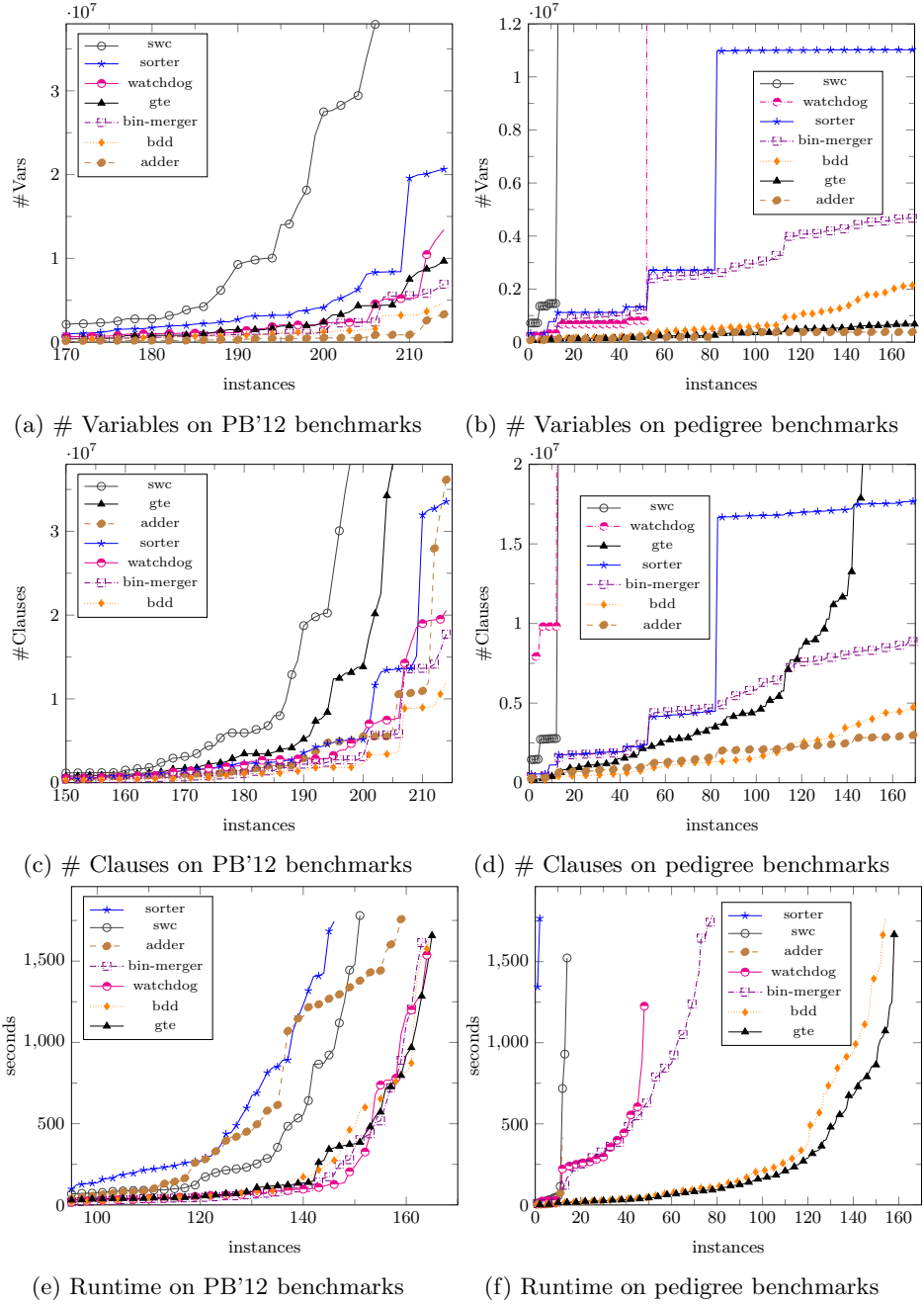
\begin{figure}
\subcaptionbox{\# Variables  on PB'12 benchmarks \label{Fi:var-pb}}[.5\textwidth]
{
\begin{tikzpicture}[scale=0.75]
\begin{axis}[xlabel=instances, ylabel=\#Vars, xmin=170, xmax=215, ymin=0, ymax=38000000,
  legend entries={ \SSZ{swc},\SSZ{sorter}, \SSZ{watchdog},\SSZ{gte}, \SSZ{bin-merger}, \SSZ{bdd},\SSZ{adder}},
minor tick num=1,
y label style={at={(axis description cs:0.1,0.5)},anchor=south},
  legend style={ legend pos=north west,}
]
 \addplot [mark repeat=2,darkgray,mark=o] table {plot-data/var-pb12/swc.dat};
 \addplot [mark repeat=2,blue,mark=star] table {plot-data/var-pb12/sorter.dat}; 
 \addplot [mark repeat=3,magenta,mark=halfcircle*] table {plot-data/var-pb12/watchdog.dat};
 \addplot [black,mark phase=1,mark repeat=2,mark=triangle*] table {plot-data/var-pb12/gen-tot.dat};
 \addplot [mark repeat=2,densely dashdotted,violet,mark=square] table {plot-data/var-pb12/bin-merge.dat};
 \addplot [mark repeat=3,densely dotted,orange,mark=diamond*] table {plot-data/var-pb12/bdd.dat};
 \addplot [mark repeat=2,dashed,brown,mark=*] table {plot-data/var-pb12/adder.dat};
\end{axis}
\end{tikzpicture}
}
\subcaptionbox{\# Variables on pedigree benchmarks \label{Fi:var-ped}}[.5\textwidth]
{
\begin{tikzpicture}[scale=0.75]
\begin{axis}[xlabel=instances, ylabel=\#Vars, xmin=0, xmax=170, ymin=0, ymax=12000000,
  legend entries={ \SSZ{swc},\SSZ{watchdog},\SSZ{sorter}, \SSZ{bin-merger}, \SSZ{bdd},\SSZ{gte}, \SSZ{adder}},
minor tick num=1,
y label style={at={(axis description cs:0.1,0.5)},anchor=south},
  legend style={ anchor=south, at={(.75,.40)}}
]
 \addplot [mark repeat=1,darkgray,mark=o] table {plot-data/var-pedigree/swc.dat};
 \addplot [mark repeat=3,mark phase=3,densely dashdotdotted,magenta,mark=halfcircle*] table {plot-data/var-pedigree/watchdog.dat};
 \addplot [mark repeat=6,blue,mark=star] table {plot-data/var-pedigree/sorter.dat}; 
 \addplot [mark repeat=6,densely dashdotted,violet,mark=square] table {plot-data/var-pedigree/bin-merge.dat};
 \addplot [mark repeat=4,densely dotted,orange,mark=diamond*] table {plot-data/var-pedigree/bdd.dat};
 \addplot [black,mark phase=2,mark repeat=4,mark=triangle*] table {plot-data/var-pedigree/gen-tot.dat};
 \addplot [mark repeat=6,dashed,brown,mark=*] table {plot-data/var-pedigree/adder.dat};
\end{axis}
\end{tikzpicture}
}
\subcaptionbox{\# Clauses  on PB'12 benchmarks \label{Fi:clause-pb}}[.5\textwidth]
{
\begin{tikzpicture}[scale=0.75]
\begin{axis}[xlabel=instances, ylabel=\#Clauses, xmin=150, xmax=215, ymin=0, ymax=38000000,
  legend entries={ \SSZ{swc},\SSZ{gte},\SSZ{adder},\SSZ{sorter}, \SSZ{watchdog}, \SSZ{bin-merger}, \SSZ{bdd},},
minor tick num=1,
y label style={at={(axis description cs:0.1,0.5)},anchor=south},
  legend style={ legend pos=north west,}
]
 \addplot [mark repeat=2,darkgray,mark=o] table {plot-data/clause-pb12/swc.dat};
 \addplot [black,mark phase=1,mark repeat=2,mark=triangle*] table {plot-data/clause-pb12/gen-tot.dat};
 \addplot [mark repeat=2,dashed,brown,mark=*] table {plot-data/clause-pb12/adder.dat};
 \addplot [mark repeat=2,blue,mark=star] table {plot-data/clause-pb12/sorter.dat}; 
 \addplot [mark repeat=3,magenta,mark=halfcircle*] table {plot-data/clause-pb12/watchdog.dat};
 \addplot [mark repeat=2,densely dashdotted,violet,mark=square] table {plot-data/clause-pb12/bin-merge.dat};
 \addplot [mark repeat=3,densely dotted,orange,mark=diamond*] table {plot-data/clause-pb12/bdd.dat};
\end{axis}
\end{tikzpicture}
}
\subcaptionbox{\# Clauses on pedigree benchmarks \label{Fi:clause-ped}}[.5\textwidth]
{
\begin{tikzpicture}[scale=0.75]
\begin{axis}[xlabel=instances, ylabel=\#Clauses, xmin=0, xmax=170, ymin=0, ymax=20000000,
  legend entries={ \SSZ{swc},\SSZ{watchdog},\SSZ{gte}, \SSZ{sorter}, \SSZ{bin-merger}, \SSZ{bdd},\SSZ{adder}},
minor tick num=1,
y label style={at={(axis description cs:0.1,0.5)},anchor=south},
  legend style={ anchor=south, at={(.30,.45)}}
]
 \addplot [mark repeat=1,darkgray,mark=o] table {plot-data/clause-pedigree/swc.dat};
 \addplot [mark repeat=3,mark phase=3,densely dashdotdotted,magenta,mark=halfcircle*] table {plot-data/clause-pedigree/watchdog.dat};
 \addplot [black,mark phase=2,mark repeat=4,mark=triangle*] table {plot-data/clause-pedigree/gen-tot.dat};
 \addplot [mark repeat=6,blue,mark=star] table {plot-data/clause-pedigree/sorter.dat}; 
 \addplot [mark repeat=6,densely dashdotted,violet,mark=square] table {plot-data/clause-pedigree/bin-merge.dat};
 \addplot [mark repeat=6,densely dotted,orange,mark=diamond*] table {plot-data/clause-pedigree/bdd.dat};
 \addplot [mark repeat=6,dashed,brown,mark=*] table {plot-data/clause-pedigree/adder.dat};
\end{axis}
\end{tikzpicture}
}
\subcaptionbox{Runtime on PB'12 benchmarks \label{Fi:run-pb}}[.5\textwidth]
{
\begin{tikzpicture}[scale=0.75]
\begin{axis}[xlabel=instances, ylabel=seconds, xmin=95, xmax=170, ymin=0,
  legend entries={ \SSZ{sorter},\SSZ{swc},\SSZ{adder},\SSZ{bin-merger}, \SSZ{watchdog},\SSZ{bdd},\SSZ{gte}},
minor tick num=1,
y label style={at={(axis description cs:0.05,0.5)},anchor=south},
  legend style={ legend pos=north west,}
]
 \addplot [mark repeat=2,blue,mark=star] table {plot-data/runtime-pb12/sorter.dat}; 
 \addplot [mark repeat=2,darkgray,mark=o] table {plot-data/runtime-pb12/swc.dat};
 \addplot [mark repeat=2,dashed,brown,mark=*] table {plot-data/runtime-pb12/adder.dat};
 \addplot [mark repeat=2,densely dashdotted,violet,mark=square] table {plot-data/runtime-pb12/bin-merge.dat};
 \addplot [mark repeat=3,magenta,mark=halfcircle*] table {plot-data/runtime-pb12/watchdog.dat};
 \addplot [mark repeat=3,densely dotted,orange,mark=diamond*] table {plot-data/runtime-pb12/bdd.dat};
 \addplot [black,mark phase=1,mark repeat=2,mark=triangle*] table {plot-data/runtime-pb12/gen-tot.dat};
\end{axis}
\end{tikzpicture}
}
\subcaptionbox{Runtime on pedigree benchmarks \label{Fi:run-ped}}[.5\textwidth]
{
\begin{tikzpicture}[scale=0.75]
\begin{axis}[xlabel=instances, ylabel=seconds, xmin=0, xmax=170, ymin=0,
  legend entries={ \SSZ{sorter},\SSZ{swc},\SSZ{adder},\SSZ{watchdog},\SSZ{bin-merger}, \SSZ{bdd},\SSZ{gte}},
minor tick num=1,
y label style={at={(axis description cs:0.05,0.5)},anchor=south},
  legend style={ anchor=south, at={(.65,.5)}}
]
 \addplot [mark repeat=1,blue,mark=star] table {plot-data/runtime-pedigree/sorter.dat}; 
 \addplot [mark repeat=1,darkgray,mark=o] table {plot-data/runtime-pedigree/swc.dat};
 \addplot [mark repeat=2,dashed,brown,mark=*] table {plot-data/runtime-pedigree/adder.dat};
 \addplot [mark repeat=3,mark phase=3,magenta,mark=halfcircle*] table {plot-data/runtime-pedigree/watchdog.dat};
 \addplot [mark repeat=4,densely dashdotted,violet,mark=square] table {plot-data/runtime-pedigree/bin-merge.dat};
 \addplot [mark repeat=4,densely dotted,orange,mark=diamond*] table {plot-data/runtime-pedigree/bdd.dat};
 \addplot [black,mark phase=2,mark repeat=4,mark=triangle*] table {plot-data/runtime-pedigree/gen-tot.dat};
\end{axis}
\end{tikzpicture}
}

\caption{Cactus plots of number of variables, number of clauses and runtimes}
\label{Fi:results}
\end{figure}

\tableref{properties} shows the characteristics of the 
benchmarks used in this evaluation. \#PB denotes the average number of 
PBCs per instance. \#lits, $k$, max $w_i$, $\sum w_i$ and \#diff $w_i$ denote the 
per constraint per instance average of input literals, bound, the largest weight, maximum
possible weighted sum and the number of distinct weights.
%
PB'12 benchmarks are a mix of 
crafted as well as 
industrial benchmarks, whereas all of the \pedigree benchmarks are 
from the same biological problem~\cite{pedigrees}. The PB'12 benchmarks 
have on average several PBCs, however, they are relatively small in
magnitude.
In contrast, the \pedigree benchmarks contain one large PB constraint with very large 
total weighted sum. \pedigree
benchmarks have only two distinct values of weights, thus making them good candidates
for using GTE.


\vspace{2mm}
\noindent \BF{Results: }
\tableref{solved} shows the number of instances solved using different 
encodings.
\sorter, \adder and \swc 
perform worse than the remaining encodings for both sets of 
benchmarks. The first two are not arc-consistent
therefore the SAT solver is not able to infer as much information as with arc-consistent 
encodings. \swc, though arc-consistent, generates a large number of auxiliary 
variables and clauses, which deteriorates the performance of the SAT solver. 

\gte provides a competitive performance to \bdd, \binmerge 
and \watchdog for 
PB'12. However, 
only the \gte and \bdd encodings are able to tackle \pedigree benchmarks,
which contain a large number of literals and only two 
different coefficients. Unlike other encodings, \gte and \bdd are able to 
exploit the characteristics of these benchmarks.

%
\swc requires significantly large number of variables as the value of $k$ increases, whereas \bdd and \gte keep
the variable explosion in check due to reuse of variables on similar
combinations (\figrefs{var-pb}{var-ped}). This reuse of auxiliary variables is even more 
evident on \pedigree benchmarks (\figref{var-ped}) as these benchmarks have only two
different coefficients resulting in low number of combinations. $k$-simplification
also helps \gte in keeping the number of variables low as all the combinations weighing
more than $k+1$ are mapped to $k+1$.

Number of clauses required for \gte is quite large as compared to some other encodings
(\figrefs{clause-pb}{clause-ped}). \gte requires clauses to be generated for all the 
combinations even though most of them produce the same value for the weighted sum, thus reusing the same variable. Though \bdd
has an exponential worst case, in practice it appears to
generate smaller formulas (\figrefs{clause-pb}{clause-ped}). 
%

\figref{run-pb} shows that \gte provides a competitive performance
with respect to \binmerge, \watchdog and \bdd. Runtime on \pedigree benchmarks as shown in
\figref{run-ped} establishes \gte as the clear winner with \bdd performing a close second.
The properties that \gte and \bdd share help them perform better on \pedigree benchmarks
as they are not affected by large magnitude of weights in the PBCs.

\section{Conclusion \label{Se:conclusion}}
\Omit{
SAT solvers are becoming increasingly efficient. Given the importance of PBCs in modeling 
real world problems, research efforts are warranted to encode PBCs into a SAT formula in a
manner that enables solving these problems efficiently.
}

Many real-world problems can be formulated using pseudo-Boolean constraints (PBC).
Given the advances in SAT technology, it becomes crucial how to encode PBC into SAT,
such that SAT solvers can efficiently solve the resulting formula.

In this paper, an arc-consistency preserving generalization of the Totalizer encoding is proposed for encoding
PBC into SAT. 
Although the proposed encoding is exponential in the worst case, the new Generalized 
Totalizer encoding (GTE) is very competitive in relation with other PBC encodings. 
Moreover, experimental results show that when the number of different weights in PBC
is small, it clearly outperforms all other encodings.
As a result, we believe the impact of GTE can be extensive, since one can further
extend it into incremental settings~\cite{martins-cp14}.


\vspace{2mm}
\noindent \BF{Acknowledgments: }
This work is partially supported by the ERC project 280053,
FCT grants AMOS (CMUP-EPB/TIC/0049/2013), POLARIS (PTDC/\-EIA-CCO/123051/2010),
and INESC-ID's multiannual PIDDAC funding UID/\-CEC/50021/2013.

\bibliographystyle{splncs03}
\bibliography{biblio}

\end{document}